\documentclass[10pt,journal,twocolumn]{IEEEtran}
\usepackage{times,epsfig,amsbsy,amssymb,float,color}
\usepackage[cmex10]{amsmath}
\usepackage{mdwmath, mathrsfs, bbm}
\usepackage{mdwtab}
\usepackage{amsfonts}
\usepackage{subfigure}
\usepackage{multicol}
\usepackage{amsmath,amssymb}
\usepackage{cases}
\usepackage{booktabs}
\usepackage{algorithm}
\usepackage{stfloats}
\usepackage{color}
\usepackage{algorithmicx}
\usepackage{algpseudocode}
\usepackage{mathtools}
\usepackage{authblk}
\algdef{SE}[DOWHILE]{Do}{doWhile}{\algorithmicdo}[1]{\algorithmicwhile\ #1}

\newtheorem{theorem}{Theorem}

\IEEEoverridecommandlockouts

\usepackage[numbers,sort&compress]{natbib}

\usepackage{flushend}

\usepackage{stackengine}
\usepackage{scalerel}
\def\clipeq{\!\mathrm{=}\!}
\def\Lla{\Longleftarrow\!\!\!\!}
\def\Lra{\!\!\!\!\Longrightarrow}
\newcommand\xleftrightarrows[3][0]{%
  \mathrel{%
  \ifthenelse{\equal{#1}{0}}%
  {\stackunder[2pt]{\stackon[3pt]{$\Lla\Lra$}%
     {$\scriptstyle#2$}}{$\scriptstyle#3$}}%
  {\stackunder[2pt]{\stackon[3pt]{$\Lla\hstretch{#1}{\clipeq}\Lra$}%
     {$\scriptstyle#2$}}{$\scriptstyle#3$}}%
  }%
}

\input epsf
\title{\huge A Physical Layer Network Coding Design for 5G Network MIMO}
\author{Tong~Peng, Yi~Wang, Alister G. Burr and Mohammad Shikh-Bahaei

\thanks{T. Peng was with the Department of Electronics, University of York, and now is with the Centre for Telecommunications Research, Department of Informatics, King's college London, UK (e-mail: tong.peng@york.ac.uk, tong.peng@kcl.ac.uk). Y. Wang and A. G. Burr are with the Department of Electronics, University of York, UK	
(e-mails: yi.wang@york.ac.uk, alister.burr@york.ac.uk). M. Shikh-Bahaei is with the Centre for Telecommunications Research, Department of Informatics, King's college London, UK (e-mail: m.sbahaei@kcl.ac.uk).

This research is funded by EPSRC NetCoM project EP/K040006/1 and partially by EPSRC IoSIRE project EP/P022723/1.}
}

\begin{document}
\maketitle\pagestyle{empty}
\begin{abstract} 
This paper presents a physical layer network coding (PNC) approach for network MIMO (N-MIMO) systems to release the heavy burden of backhaul load. The proposed PNC approach is applied for uplink scenario in binary systems, and the design guideline serves multiple mobile terminals (MTs) and guarantees unambiguous recovery of the message from each MT. We present a novel PNC design criterion first based on binary matrix theories, followed by an adaptive optimal mapping selection algorithm based on the proposed design criterion. In order to reduce the real-time computational complexity, a two-stage search algorithm for the optimal binary PNC mapping matrix is developed. Numerical results show that the proposed scheme achieves lower outage probability with reduced backhaul load compared to practical CoMP schemes which quantize the estimated symbols from a log-likelihood ratio (LLR) based multiuser detector into binary bits at each access point (AP).

\end{abstract}


\section{Introduction}
\label{sec:introduction}

The concept of network multiple input, multiple output (N-MIMO) \cite{M.V.Clark} has been known for some time as a means to overcome the inter-cell interference in fifth generation (5G) dense cellular networks, by allowing multiple access points (APs) to cooperate to serve multiple mobile terminals (MTs). This was implemented in the coordinated multipoint (CoMP) approach standardized in LTE-A \cite{D.Lee}. More recently the Cloud Radio Access Network (C-RAN) concept has been proposed, which has similar goals \cite{R.Irmer}. However these approaches result in large loads on the backhaul network (also referred to as fronthaul in C-RAN) between APs and the central processing unit (CPU), many times the total user data rate. 


While there has been previous work addressing backhaul load reduction in CoMP and C-RAN, using, for example, Wyner-Ziv compression \cite{re-bkload0} or iterative interference cancellation \cite{re-bkload2}, the resulting total backhaul load remains typically several times the total user data rate. A novel approach was introduced in \cite{Zhang2006}, based on physical layer network coding (PNC), which reduces the total backhaul load to be equal to the total user data rate. However most previous research on PNC focused on two-way relay channel (TWRC) or lattice code-based design. In \cite{Zhang2006}, a PNC based on BPSK scheme is designed for the TWRC. Compute-and-forward approach which generalizes PNC of TWRC to multiuser relay networks by utilizing lattice network coding \cite{FengLNC} is presented in \cite{CF_Nazer}, and heterogeneous modulation schemes which lead to precoded PNC design are studied in \cite{Hete_Mod}. Work in \cite{DongF2} provides a PNC design in N-MIMO system but the design focused on lower-order modulation schemes. 



In this paper, we present an adaptive PNC design with unambiguous detection of messages from all MTs to reduce the backhaul load in binary N-MIMO systems. The main contributions are listed as follows

\begin{enumerate}
\item \textbf{Novel PNC design criterion}: Different from the previous work on symbol-level PNC approaches, we focus on the bit-level PNC design that can be applied with conventional $2^m$-ary digital modulation. When QAM modulation schemes are used at MTs, PNC has to solve the so-called singular fading problem which typically unavoidable at the multiple access stage under some circumstances at each AP. Failure to resolve such problem results in network performance degradation. In order to resolve the singular fading problem, we present a novel design criterion for PNC with unambiguous recovery capability for messages from all MTs. In addition, the proposed PNC design criterion to encode PNC directly at the bit level which allows the APs to operate over a binary field for practical application.

\item \textbf{Adaptive optimal mapping selection algorithm}: Based on the proposed PNC design criterion, a two-stage adaptive optimal mapping matrix selection algorithm is developed to achieve reduced real-time computational complexity. In the first stage, an exhaustive search among all mapping matrix candidates is required to find out the optimal mapping matrix candidate to resolve each singular fading. This exhaustive search needs to be done only once for different QAM modulation schemes which contains the majority computational complexity. The selected mapping matrix candidates are employed in the second stage during the real-time uplink period. Due a considerable reduced number of mapping matrix candidates, the real-time search algorithm can be applied in practical systems.   
\end{enumerate}

In summary, the main contribution of this paper is that a novel PNC design for binary systems with multiple MTs using conventional $2^m$ digital modulation to reduce the backhaul load in upload scenario is proposed. Unambiguous recovery of messages from all MTs is guaranteed by the novel design criterion and based on that, a two-stage adaptive mapping matrix selection algorithm is developed with reduced real-time computational complexity. The proposed mapping selection algorithm operates over binary systems in the uplink scenario and hence can be applied in the current mobile communication networks. 

The rest of the paper is organised as follows. Section II describes the N-MIMO system model and Section III presents the design criterion of the proposed PNC approach followed by a two-stage optimal mapping matrix selection algorithm. In Section IV, comparisons among the proposed PNC approach and benchmarks such as ideal CoMP with unlimited backhaul load and non-ideal CoMP schemes with limited backhaul load are given and contributions of this paper is concluded in Section V.

\section{System Model}

\begin{figure}[t]
  \centering
\begin{minipage}[t]{1\linewidth}
\centering
  \includegraphics[width=1\textwidth]{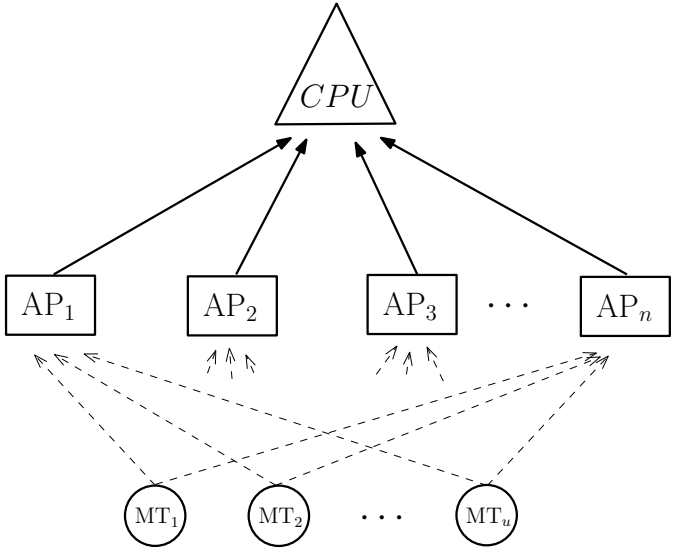}
\end{minipage}
\caption{\small The uplink of N-MIMO system with multiple MTs and APs.} \label{fig:system}
\end{figure}

The N-MIMO system model we consider in this paper is shown in Fig. \ref{fig:system}, where $u$ MTs transmit their messages to the CPU via $n$ APs. Single antenna is employed at each node. The uplink process is divided into two stages which are named as multiple access (MA) stage and backhaul (BH) stage. In the MA stage, $u$ MTs transmit the QAM modulated signals to the APs simultaneously, mathematically given by
\begin{align}
y_j = \sum_{l=1}^{u}h_{j,l}s_l + z_j, ~\text{for}~ j=1,2,\cdots,n,
\end{align} 
where $h_{j,l}$ represents the channel fading coefficient between the ${l}^{\mathrm{th}}$ MT and the $j^{\mathrm{th}}$ AP, which is assumed to be independent and identically distributed (i.i.d.) zero mean complex Gaissuan random variable and unit variance, such as Rayleigh flat fading channel coefficient. $z_j$ stands for the additive complex Gaussian noise with zero mean and variance $\sigma^2$. $s_l \triangleq \mathscr{M}(\mathbf{w}_l)$ is the modulated signal at the $l^{\mathrm{th}}$ MT, where $\mathscr{M}:\mathbb{F}_{2^m} \longrightarrow \Omega$ denotes a $2^m$-QAM modulation scheme and $m$ is the modulation order. $\Omega$ is the set of $2^m$ possible complex constellation points and $\mathbf{w}_l=[w_{l}^{(1)},\cdots,w_{l}^{(m)}]$ is the $1 \times m$ binary message vector at the $l^{\mathrm{th}}$ MT. 

At each AP, binary mapping matrices are utilised to map the superimposed signal to the network coded vector (NCV). Define a set $\mathbf{W} \triangleq [\mathbf{w}^{(1)}, \cdots, \mathbf{w}^{(u^{mu})}]$ which contains all possible combinations of joint binary messages from $u$ MTs, by multiplying an $m \times mu$ binary mapping function $\mathbf{G}_j$ at the $j^{\mathrm{th}}$ AP, the PNC encoding is then given by
\begin{align}\label{equ:pnc_encode}
\mathbf{x}_j = \mathbf{G}_j \otimes \mathbf{w}^{(p)}, ~\text{for}~ p=1,2,\cdots,u^{mu},
\end{align}
where $\mathbf{x}_j$ denotes the $m \times 1$ NCV, $\mathbf{w}^{(p)}$ denotes the $p^{\mathrm{th}}$ joint message vector with a size of $mu \times 1$ in $\mathbf{W}$, and $\otimes$ denotes the multiplication over $\mathbb{F}_2$.   

After obtain the joint message set $\mathbf{W}$, the corresponding modulated symbol combination defined by a set $\mathbf{S} \triangleq \mathscr{M}(\mathbf{W})=[\mathbf{s}^{(1)}_c, \cdots, \mathbf{s}^{(u^{mu})}_c]$ can be obtained and each element in $\mathbf{S}$ contains $u$ modulated symbols. Given a $1 \times u$ channel coefficient vector $\mathbf{h}_j\triangleq [h_{j,1}, h_{j,2}, \cdots, h_{j,u}]$ at the $j^{\mathrm{th}}$ AP, the set containing all $u^{mu}$ possible superimposed signals can be calculated by
\begin{align}\label{equ:sup_sig_vet}
\mathbf{s}_{j,\mathrm{sc}} \triangleq [s^{(1)}_{j,\mathrm{sc}}, s^{(2)}_{j,\mathrm{sc}}, \cdots, s^{(u^{mu})}_{j,\mathrm{sc}}] = \mathbf{h}_j \mathbf{S}, ~\mathrm{for}~ j = 1,\cdots, n.
\end{align}
In order to obtain $\mathbf{x}_j$ from $\mathbf{s}_{j,\mathrm{sc}}$, an estimator calculates the conditional probability of each possible NCV given a mapping matrix $\mathbf{G}_j$ and the channel coefficients $\mathbf{h}_j$ is employed at each AP. The estimator returns the log-likelihood ratio (LLR) of each bit of $\mathbf{x}_j$ which is then applied to a soft decision decoder. 

Then the NCV obtained at each AP will be forwarded to the CPU via a backhaul network in BH stage and the CPU recovers the original messages from each MT by concatenating the NCVs and multiplying by the inverse of the global PNC mapping matrix (formed by the concatenation of $\mathbf{G}_j$s used at each AP), mathematically given by
\begin{align}
\hat{\mathbf{w}} = \mathbf{G}^{-1} \otimes \mathbf{x} = ([\mathbf{G}_1 \cdots \mathbf{G}_n]^T)^{-1} \otimes [\mathbf{x}_1 \cdots \mathbf{x}_n]^T, \label{eq:PNCdecod}
\end{align}
where $\hat{\mathbf{w}}$ stands for the estimated original message vector, $\mathbf{G}$ denotes the $mu \times mu$ global mapping matrix and $\mathbf{x}$ stands for the $mu \times 1$ NCV concatenated by the NCVs from all APs. In this paper, the links in MA stage are modelled as wireless links in order to fulfil the request of 5G systems; while the backhaul links are deployed on a lossless but capacity limited 'bit-pipe'.

\section{Adaptive Binary PNC Design}

In this section, we present the novel PNC design criterion followed by a two-stage optimal mapping selection algorithm for N-MIMO systems with multiple MTs and APs. We define and study the so-called singular fading problem first, followed by proposing a method to resolve this problem based on the proposed design criterion, and finally develop a two-stage binary mapping selection algorithm with eliminated computational complexity. The detailed mathematical proof is also provided in this section to support our design criterion. 

\subsection{Singular Fading in MA Stage}
Singular fading in the MA stage is a serious problem which degrades network performance due to the undistinguished superimposed symbol groups under some conditions. We give a simple example here with $2$ MTs to define the singular fading problem and summarise the PNC design criterion after. It is worth to mention the design criterion can be extended to a general N-MIMO system with more than $2$ MTs. A discussion is given at the end of this subsection. 

The singular fading is defined as a situation in which different pairs of transmitted symbols cannot be distinguished at the receiver, mathematically given by
\begin{align}\label{equ:sfs}
h_{j,1}s_1 + h_{j,2}s_2 = h_{j,1}s^\prime_1 + h_{j,2}s^\prime_2,
\end{align}
where $s_i$ and $s^\prime_i$ stand for the QAM modulated signals at the $i^{\mathrm{th}}$ MT for $i=1,2$, and $s_i \neq s^\prime_i$. The special channel coefficient vector $\mathbf{h}_\mathrm{sf}\triangleq [h_{j,1}, h_{j,2}]$ is defined as a singular fade state (SFS). In this case, the failure in detection of $[s_1~ s_2]$ and $[s^\prime_1~ s^\prime_2]$ due to the superimposed symbols coincide in the constellation will degrade the network performance. Please note that the solution of (\ref{equ:sfs}) is not unique, hence there is more than one SFSs for each QAM scheme. 

We can calculate the values of all SFSs by substituting all possible modulated symbol combinations to (\ref{equ:sfs}), which is given by 
\begin{align}\label{eq:SFSt}
v^{(q)}_{SFS} = \frac{h_{1,2}}{h_{1,1}} = \frac{s^{(\tau)}_1-s^{(\tau^{\prime})}_1}{s^{(\tau^{\prime})}_2-s^{(\tau)}_2}, ~~\forall s^{(\tau)}_l,s^{(\tau^{\prime})}_l \in \Omega, 
\end{align}
where $v^{(q)}_{SFS}$ is an element in the SFS set $\mathbf{v}_{SFS} \triangleq [v^{(1)}_{SFS}, \cdots , v^{(L)}_{SFS}]$ for a QAM modulation scheme and the value of $L$ is different for different modulation schemes.

According to (\ref{equ:sfs}), an SFS results in ambiguous detection when different symbol combinations are transmitted from MTs. We define a \emph{clash} which contains the superimposed symbols with the same value. In order to achieve unambiguous detection, the superimposed symbols in a clash should be mapped to the same NCV. We extend the definition of clash to \emph{cluster} in which the superimposed symbols are mapped to the same NCV to reduce the decoding complexity. In this case, a cluster may contain a set of clashed symbols and individual non-clashed symbols. Then by mapping the superimposed symbols in a cluster to the same NCV, the problem of SFS is solved. Additionally the optimal PNC mapping matrix should keep different NCVs as far apart as possible in order to achieve the unambiguous detection of all messages from the MTs. 

By defining $d_{\mathrm{min}}$ as the minimum Euclidean distance between different clusters, the optimal PNC mapping matrix design criterion is given by
\begin{align}\label{eq:desgcri}
& \mathbf{x}^{(opt)}_j = \mathbf{G}^{(opt)}_j \otimes \mathbf{w}_{\mathrm{cluster}}^{(i,k)}, ~\forall \mathbf{w}_{\mathrm{cluster}}^{(i,k)} \in \mathbf{W}^{(k)}_{\mathrm{cluster}}, \\
& \text{s.t.} ~\max\limits_{\mathbf{G}_j} d_{\mathrm{min}}, ~\text{for}~ i,k=1,2,\cdots,mu, \notag 
\end{align}
where
\begin{align}\label{eq:dmint}
d_{\mathrm{min}} = \min_{\Theta(s_{j,sc}^{(\tau)}) \neq \Theta(s_{j,sc}^{(\tau^{\prime})}) }  |s_{j,sc}^{(\tau)} - s_{j,sc}^{(\tau^{\prime})}|^2, 
\end{align}
\begin{align}
\mathbf{x}^{(\tau)}_j = \Theta(s_{j,sc}^{(\tau)})=\mathbf{G}_j \otimes \mathbf{w}_{\mathrm{cluster}}^{(\tau)}, \label{eq:NCVcal}
\end{align}
\begin{align}\label{eq:Ssc}
& s^{(\tau)}_{j,sc} = s^{(\tau)}_1 + v^{(q)}_{SFS}s^{(\tau)}_2, ~\forall s^{(\tau)}_1,s^{(\tau)}_2 \in \Omega, \\ 
\forall s^{(\tau)}_{j,sc} &\in \mathbf{c}^{(\tau)}, ~\forall s^{(\tau^\prime)}_{j,sc} \in \mathbf{c}^{(\tau^\prime)}, 
~q=1,2,\cdots, L, \notag 
\end{align}
where $\mathbf{W}^{(k)}_{\mathrm{cluster}}$ denotes the binary joint message set of the $k^{\mathrm{th}}$ cluster and $\mathbf{w}_{\mathrm{cluster}}^{(i,k)}$ denotes the $i^{\mathrm{th}}$ joint message vector in $\mathbf{W}^{(k)}_{\mathrm{cluster}}$. $s_{j,sc}^{(\tau)}$ and $s_{j,sc}^{(\tau^{\prime})}$ denote for the superimposed symbols in cluster $\mathbf{c}^{(\tau)}$ and $\mathbf{c}^{(\tau^\prime)}$, respectively, and $\Theta$ denotes the PNC mapping function.

The key elements in the above ceiteria is the values of SFS. In the $2$-MT case, (\ref{eq:SFSt}) can be utilised to determine all possible SFSs for a QAM modulation scheme. However, in multiple-MT ($u>2$) case, (\ref{eq:SFSt}) is no longer suitable for mathematical representation of SFSs because the increased number of MTs in MA stage results in non-linear relationship between modulated symbols and SFS values. This is still an open question for PNC design. In this paper, we utilise $2$-MT case as the basis to cope with the multiple-MT case. For example, when $3$ MTs are served by an AP, we use (\ref{eq:SFSt}) to determine the SFS between $\text{MT}_1$ and $\text{MT}_2$ and the received symbol from $\text{MT}_3$ is treated as additional noise. Then we pair $\text{MT}_1$ (or $\text{MT}_2$) and $\text{MT}_3$ for PNC encoding and treat signals from $\text{MT}_2$ (or $\text{MT}_1$) as additional noise. We utilise received power as threshold for pairing different MTs and have been discussed in \cite{Acs}.

\subsection{Algebraic Work for Design Criterion}
In this subsection, we illustrate mathematical work to support the proposed PNC design criterion. As mentioned in the previous subsection, we need to carefully design each $\mathbf{G}_j$, $j=1,2,\cdots,n$, so that $\mathbf{G} = [\mathbf{G}_1,\cdots,\mathbf{G}_n]^T$ includes a number of row coefficients which forms
\begin{theorem} \label{theorem:full.rank}
Assuming $\mathbf{G}=G_{n\times n}(R)$, where the coefficients are from a commutative ring $R$. Source messages are drawn from a subset of $R$ and all source messages can be unambiguously decoded at the destination if and only if the determinant of the transfer matrix is a unit in $R$,
\begin{align}
\mathrm{det}(\mathbf{G}) = \mathcal{U}(R). \label{equ:det.unit}
\end{align}
\end{theorem}

\begin{IEEEproof} We first prove that (\ref{equ:det.unit}) gives the sufficient and necessary conditions that make a matrix $\mathbf{B}$ invertible in N-MIMO networks. Suppose $\mathbf{B}$ is invertible, then there exists a matrix $\mathbf{C}\in G_{n\times n}(R)$ such that $\mathbf{B}\mathbf{C} = \mathbf{C}\mathbf{B} = \mathbf{I}_n$. This implies $1 = \mathrm{det}(\mathbf{I}_n) = \mathrm{det}(\mathbf{B}\mathbf{C}) = \mathrm{det}(\mathbf{B})\mathrm{det}(\mathbf{C})$. According to the definition of a unit, we say $\mathrm{det}(\mathbf{B})\in U(R)$.

We know $\mathbf{B}\cdot \mathrm{adj}(\mathbf{B}) = \mathrm{adj}(\mathbf{B})\cdot \mathbf{B} = \mathrm{det}(\mathbf{B})\mathbf{I}_n$. If $\mathrm{det}(\mathbf{B})\in U(R)$, we have
\begin{align}
\mathbf{B} \cdot (\mathrm{det}(\mathbf{B})^{-1}\mathrm{adj}(\mathbf{B}) ) &=   (\mathrm{det}(\mathbf{B})^{-1}\mathrm{adj}(\mathbf{B}) ) \mathbf{B}  \notag\\
&= \mathrm{det}(\mathbf{B})^{-1} \mathrm{det}(\mathbf{B}) = \mathrm{I}_n.
\end{align}
Hence, $\mathbf{C} = (\mathrm{det}(\mathbf{B})^{-1}\mathrm{adj}(\mathbf{B}) )$ is the inverse of $\mathbf{B}$ since $\mathbf{B}\mathbf{C} = \mathbf{C}\mathbf{B} = \mathbf{I}_n$.

If $\mathbf{B}$ is invertible, then its inverse $\mathbf{B}^{-1}$ is uniquely determined. Assuming $\mathbf{B}$ has two inverses, say, $\mathbf{C}$ and $\mathbf{C}^{\prime}$. Then
\begin{align}
\mathbf{B}\cdot \mathbf{C} &= \mathbf{C}\cdot \mathbf{B} = \mathbf{I}_n, \\
\mathbf{B}\cdot \mathbf{C}^{\prime} &= \mathbf{C}^{\prime}\cdot \mathbf{B} = \mathbf{I}_n,
\end{align}
hence we have
\begin{align}
\mathbf{C} = \mathbf{C}\cdot \mathbf{I}_n = \mathbf{C}\cdot \mathbf{B}\cdot \mathbf{C}^{\prime} = \mathbf{I}_n\cdot \mathbf{C}^{\prime} = \mathbf{C}^{\prime}.
\end{align}
It proves the uniqueness of the invertible matrix $\mathbf{B}$ over $R$.

Assume $\mathbf{a}\neq \mathbf{a^{\prime}}$, $\mathbf{B}\cdot \mathbf{a} = \mathbf{F}$, $\mathbf{B}\cdot \mathbf{a}^{\prime} = \mathbf{F^{\prime}}$, and $\mathbf{F} = \mathbf{F^{\prime}}$. This means
\begin{align}
\mathbf{a} = \mathbf{B}^{-1}\cdot \mathbf{F} = \mathbf{B}^{-1}\cdot \mathbf{F^{\prime}} = \mathbf{a^{\prime}}.
\end{align}
This contradicts $\mathbf{a}\neq \mathbf{a^{\prime}}$. Hence it ensures unambiguously decodability:
\begin{equation}
\mathbf{B}\cdot \mathbf{a} \neq \mathbf{B}\cdot \mathbf{a^{\prime}}, ~ \forall\mathbf{a}\neq \mathbf{a^{\prime}}.
\end{equation}
\end{IEEEproof}

Let $I_{\nu}(G_{m\times n}(R))$ denotes the ideal in $R$ generated by all $\nu\times\nu$ minors of $G_{m\times n}(R)$, where $\nu=1,2,\cdots,r=\min\{m,n\}$. A $\nu\times \nu$ minor of $G_{m\times n}(R)$ is the determinant of a $\nu\times\nu$ matrix obtained by deleting $m-\nu$ rows and $n-\nu$ columns. Hence there are $\binom{m}{\nu}\binom{n}{\nu}$ minors of size $\nu\times\nu$. $I_{\nu}(G_{m\times n}(R))$ is the ideal of $R$ generated by all these minors.

\textit{Design criterion}: The destination is able to unambiguously decode $u$ source messages if:
\begin{enumerate}
\item $u\geq\max\left\{\nu\mid \mathrm{Ann}_R(I_{\nu}(\mathbf{G}_j)) = \langle 0 \rangle \right\}$, $\forall j=1,2,\cdots,n$,
\item $\mathbf{G}_j = \arg\max\limits_{\mathbf{M}_j} \left\{ I\left( \overrightarrow{Y}; \overrightarrow{F}_j \right)  \right\}$,
\end{enumerate}
where $\langle x\rangle$ denotes the ideal generated by $x$. Condition $1$ can be proved as follows. According to Laplace's theorem, every $(\nu+1)\times (\nu+1)$ minor of $G_{m\times n}(R)$ must lie in $I_{\nu}(G_{m\times n}(R))$. This suggests an ascending chain of ideals in $R$:
\begin{align}
\langle 0 \rangle = I_{r+1}(\mathbf{G}_j)&\subseteq I_{r}(\mathbf{G}_j)\subseteq\cdots\subseteq I_{1}(\mathbf{G}_j)\notag\\
&\subseteq I_{0}(\mathbf{G}_j) =R. \label{equ:chain}
\end{align}
Computing the annihilator of each ideal in (\ref{equ:chain}) produces another ascending chain of ideals,
\begin{align}
\langle 0 \rangle =\mathrm{Ann}_R(R)&\subseteq\mathrm{Ann}_R(I_{1}(\mathbf{G}_j))\subseteq\cdots\subseteq\mathrm{Ann}_R(I_{r}(\mathbf{G}_j)) \notag\\
&\subseteq\mathrm{Ann}_R(\langle 0 \rangle ) =R.
\end{align}
It is obvious that:
\begin{align}
&\mathrm{Ann}_R(I_{k}(\mathbf{G}_j))\neq \langle 0 \rangle \notag \\
\Rightarrow & \mathrm{Ann}_R(I_{k^\prime}(\mathbf{G}_j))\neq \langle 0 \rangle, ~~\forall k\leq k^\prime.
\end{align}

The maximum value of $\nu$ which satisfies $\mathrm{Ann}_R(I_{\nu}(\mathbf{G}_j))= \langle 0 \rangle$ guarantees that $I_{k}(\mathbf{G}_j)\in R$, $\forall k<\nu$. Hence we define the rank of $\mathbf{G}_j$ as $\mathrm{rk}(\mathbf{G}_j)=\max\left\{\nu\mid \mathrm{Ann}_R(I_{\nu}(\mathbf{G}_j)) = \langle 0 \rangle \right\}$. Suppose that $\mathbf{G}_k\in G_{m\times p}(R)$ and $\mathbf{G}_{k^\prime}\in G_{p\times n}(R)$, then $\mathrm{rk}(\mathbf{G}_k\mathbf{G}_{k^\prime})\leq \min\{\mathrm{rk}(\mathbf{G}_k),\mathrm{rk}(\mathbf{G}_{k^\prime})\}$, and we can easily prove that $0\leq \mathrm{rk}(G_{m\times n}(R))\leq \min\{m,n\}$. Thus, in order to guarantee there are at least $u$ unambiguous linear equations available at the CPU, $\mathrm{rk}(\mathbf{G}_j)$ must be at least $u$, $\forall j=1,2,\cdots,n$.

The special case of condition $1$ is that the entry of the coefficient matrix $\mathbf{G}_j\in G_{m\times n}(F)$ is from a finite field $F\in\mathbb{F}$. Then condition $1$ may be changed to \textquotedblleft the maximum number of linearly independent rows (or columns)\textquotedblright~ since $\mathrm{Ann}_R(I_{\nu}(\mathbf{G}_j))= \langle 0 \rangle$ if and only if $I_{\nu}(\mathbf{G}_j)\neq 0$. In other words, the largest $\nu$ such that the $\nu\times \nu$ minor of $\mathbf{G}_j$ is a non-zero divisor represents how many reliable linear combinations the $j^{\mathrm{th}}$ layer may produce. Hence condition $1$ is a strict definition which ensures unambiguous decodability of the $u$ sources. Condition $3$ ensures that the selected coefficient matrix maximises the mutual information of the particular layer, giving finally the maximum overall throughput.

\subsection{Binary Adaptive Mapping Selection Algorithm}

In this section, we present a binary adaptive mapping selection (BMAS) algorithm based on the design criterion presented in the previous subsection. Like the work in \cite{DongF2}, our proposed algorithm comprises two procedures, the first of which is an Off-line search and the second is an On-line search. 

Before the proposed Off-line search, the values of SFS for QAM modulation schemes can be determined by using (\ref{eq:SFSt}). Then we define an $L \times u_j$ matrix $\mathbf{H}_\mathrm{SFS}=[\mathbf{h}_{\mathrm{sfs}_1}, \cdots, \mathbf{h}_{\mathrm{sfs}_L}]^T$ contains all SFS for a QAM modulation scheme, where $\mathbf{h}_{\mathrm{sfs}_q} = [1 ~v^{(q)}_{SFS}]$ for $q=1,\cdots, L$. By substituting all $v_{SFS}$s into (\ref{eq:Ssc}), the superimposed symbols can be obtained. According to the design criterion, the mapping matrices used at each AP should encode the superimposed constellation points within one cluster to the same NCV and additionally, the Euclidean distance between different NCVs should be maximised. In that case, an exhaustive search among all $m \times mu$ binary matrices is implemented in the proposed Off-line search to calculate all possible NCVs using (\ref{eq:NCVcal}) and hence the minimum Euclidean distance between different clusters can be obtained by using (\ref{eq:dmint}). The the optimal mapping matrix candidate for each SFS are selected according to the design criterion (\ref{eq:desgcri}) and stored at the APs and CPU. During the real-time transmission, an On-line search algorithm is proposed among the matrix candidates output from the Off-line search and finally, an invertible global mapping matrix with maximum value of $d_{\mathrm{min}}$ will be selected. 

An potential issue of the proposed BMAS algorithm is that the large number of SFSs in QAM schemes with higher modulation orders increases the computational complexity and searching latency in the proposed On-line search. For example, a search among $389$ $4 \times 8$ binary matrices for $16$QAM to resolve all SFSs puts the proposed On-line search algorithm in real-time application to a serious trouble. However, according to our research we found that many different SFSs generate the same clashes so that they could be resolved by the same binary matrices. Furthermore, we have found that the appearance probability of each SFS is different and we can ignore those \textquotedblleft nonactive\textquotedblright ~SFSs with low appearance probabilities (which refers to the statistical possibility for different superimposed symbols dropped in a clash is lower than $10^{-5}\%$) to reduce the number of mapping matrices utilised in the proposed On-line search with eliminated performance degradation.

\section{Numerical Results}
In this section, we evaluate the outage probability ($P_{\mathrm{out}}$) performances of the proposed BMAS algorithm and present comparisons to ideal and non-ideal CoMP over different N-MIMO networks. $4$QAM and $16$QAM modulation schemes are employed in the simulation. Convolution code with rate of $2/3$ is utilised and it can be replaced by more powerful coding schemes. In ideal CoMP, the bandwidth of backhaul network is assumed unlimited so that a joint multiuser ML detection is employed. While in non-ideal CoMP, bandwidth-limited backhaul network is considered and a $2$-bit/$4$-bit quantizer which quantizes the estimated symbols from an LLR based multiuser detection into binary bits is employed. We illustrate the $P_{\mathrm{out}}$ performance comparison between different approaches in a $5$-node network ($2$MT-$2$AP-CPU) and then extend the network to $6$-node ($3$MT-$2$AP-CPU) and $7$-node ($3$MT-$3$AP-CPU) scenarios. 

\begin{figure}[t]
  \centering
\begin{minipage}[t]{1\linewidth}
\centering
  \includegraphics[width=1\textwidth]{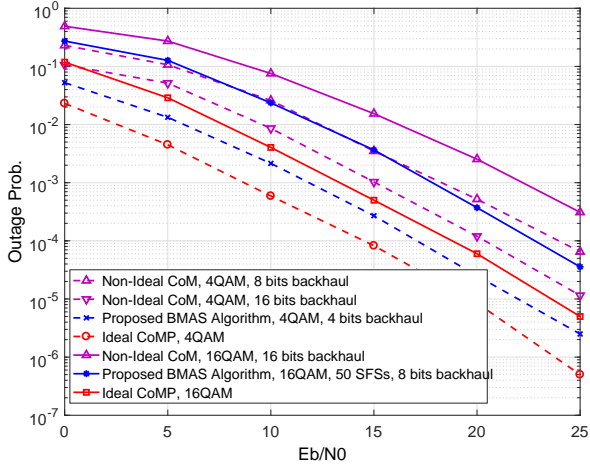}
\end{minipage}
\caption{\small The outage probability performance of $2$MT-$2$AP-CPU network, $4$QAM at MTs.} \label{fig:2MT2AP}
\end{figure}
Fig. \ref{fig:2MT2AP} illustrates $P_{\mathrm{out}}$ performance comparison among different schemes over a $2$MT-$2$AP-CPU N-MIMO network. Single antenna is employed at each node. We note that the ideal CoMP achieves the optimal $P_{\mathrm{out}}$ due to the joint ML detection with unlimited backhaul. While a performance degradation is observed in non-ideal CoMP when quantizers with limited bits are utilised to release the high backhaul load burden. As illustrated in the figure, the proposed BMAS approach is superior to the non-ideal CoMP with reduced backhaul load, even when an increased quantisation bits is used. Also the proposed BMAS approach achieves the same diversity order as ideal and non-ideal CoMP. When $16$QAM is employed at each MT, degradation of the $P_{\mathrm{out}}$ performances are observed from the figure. In the proposed BMAS algorithm, $50$ most active SFSs are utilised and according to comparison in the figure, the proposed algorithm outperforms the non-ideal CoMP with half backhaul load.

\begin{figure}[t]
  \centering
\begin{minipage}[t]{1\linewidth}
\centering
  \includegraphics[width=1\textwidth]{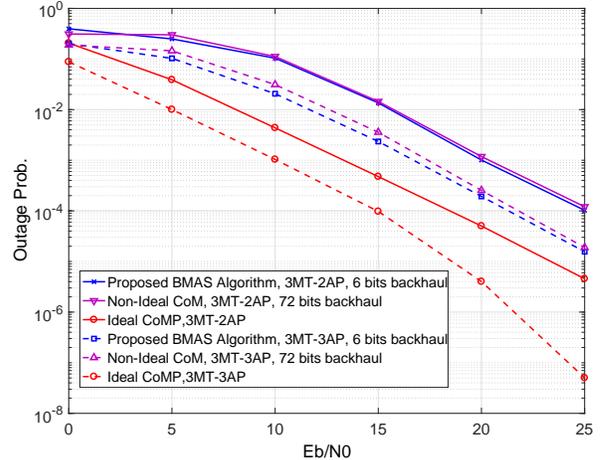}
\end{minipage}
\caption{\small The outage probability performance of $3$MT-$2$AP-CPU and $3$MT-$3$AP-CPU network.} \label{fig:3MT2/3AP}
\end{figure}
In Fig. \ref{fig:3MT2/3AP}, the outage probability performances of different schemes in $3$MT-$2$AP-CPU and $3$MT-$3$AP-CPU networks are given. In the networks with $3$ MTs, an increased backhaul load ($6$ bits in BMAS and $72$ bits in non-ideal CoMP with $4$-bit quantiser) is required. Similar to the performances in Fig. \ref{fig:2MT2AP}, ideal CoMP achieves the optimal $P_{\mathrm{out}}$ in both networks and the proposed BMAS approach is superior to the non-ideal CoMP with much less backhaul load.

\section{Future Work and Conclusion}
In this paper, we present a novel PNC design criterion for
binary systems and based on this criterion, a two-stage optimal
mapping matrix selection algorithm with low computational
complexity is developed for the uplink of N-MIMO networks
in order to reduce backhaul load. The proposed design criterion
are supported by the theorems developed and guarantees
unambiguous detection of the messages from all MTs at the
CPU. The proposed BMAS algorithm is divided into two parts
in order to reduce the real-time computational complexity
for practical implementation. Comparing
with the non-ideal CoMP, the proposed algorithm achieves
lower outage probability with a reduced traffic in backhaul
networks.

The practical application of the PNC designs mentioned in Section I and the proposed algorithm could be extended in the following areas. A study of applying the proposed design criterion and search algorithm in binary systems with full-duplex APs \cite{FD-D2D} has been started. Practical PNC design with spectrum optimisation in \cite{CLD-D2D}-\cite{CLD-D2D_1} provide another research direction and how to apply PNC with the retransmission schemes, such as ARQ and HARQ techniques \cite{COP-ARQ, FD-ARQ}, has attracted the researchers' attention. Moreover, the PNC application with multimedia symbols \cite{video-WLAN} is required in order to serve the 5G systems and achieve massive data transmissions with high accuracy and low latency.

\small
\bibliographystyle{IEEEtran}

\end{document}